\documentclass[fleqn,usenatbib]{mnras}

\usepackage{newtxtext,newtxmath}

\usepackage[T1]{fontenc}
\usepackage{ae,aecompl}

\usepackage{graphicx}	
\usepackage{amsmath}	
\usepackage{changes}
\usepackage{relsize}

\def\apj{ApJ}
\def\apjl{ApJL}
\def\apjs{ApJ Suppl. Ser.}

\def\apspr{Astroph. Sp. Phys. Rev.}
\def\aap{A\&A}

\def\mnras{MNRAS}

\def\beq#1{\begin{equation}\label{#1}}
\def\eeq{\end{equation}}
\def\beqa#1{\begin{eqnarray}\label{#1}}
\def\eeqa{\end{eqnarray}}

\def\Eq#1{Eq.~(\ref{#1})} 

\def\myfrac#1#2{\left(\frac{#1}{#2}\right)}

\def\comment#1{\relax}

\title[Orbital eccentricity and period change in SS433]{Discovery of orbital Eccentricity and Evidence for orbital Period Increase of SS433}

\author[A.M. Cherepashchuk et al.]{
A.M. Cherepashchuk$^{1}$\thanks{E-mail: cherepashchuk@gmail.com},
A.A. Belinski$^1$\thanks{E-mail: aleks@sai.msu.ru},
A.V. Dodin$^1$\thanks{E-mail: dodin\_nv@mail.ru} and
K.A. Postnov$^{1,2}$\thanks{E-mail: pk@sai.msu.ru}
\\
$^{1}$ Sternberg Astronomical Institute, M.V. Lomonosov Moscow State University, 13, Universitetskij pr., 119234, Moscow, Russia\\
$^{2}$ Kazan Federal University, Kremlevskaya 18, 420008 Kazan, Russia
}

\date{Accepted XXX. Received YYY; in original form ZZZ}

\pubyear{2021}

\begin{document}
\label{firstpage}
\pagerange{\pageref{firstpage}--\pageref{lastpage}}
\maketitle

\begin{abstract}
The examination of long-term (1979-2020) photometric observations of SS433 enabled us to discover  a non-zero orbital eccentricity of $e=0.05\pm 0.01.$ We have also found evidence for a secular increase in the orbital period at a rate of $\dot P_\mathrm{b}=(1.0\pm0.3)\times10^{-7}$~s~s$^{-1}$. The binary orbital period increase rate makes it possible to improve the estimate of the binary mass ratio $q=M_\mathrm{X}/M_\mathrm{V}>0.8$, where $M_\mathrm{X}$ and $M_\mathrm{V}$ are the masses of the relativistic object and the optical star, respectively. For an optical star mass of 10${\rm M}_\odot$, the mass of the relativistic object (a black hole) is $M_\mathrm{X}>8{\rm M}_\odot$. A neutron star in SS433 is reliably excluded because in that case the orbital period should decrease, in contradiction to observations. The derived value of $\dot P_\mathrm{b}$ sets a lower limit on the mass-loss rate in the Jeans mode from the binary system $\gtrsim 7\times 10^{-6} {\rm M}_\odot\mathrm{yr}^{-1}$.  The discovered orbital ellipticity of SS433 is consistent with the model of the slaved accretion disc tracing the precession of the misaligned optical star's rotational axis.

\end{abstract}




\section{Introduction}

The unique Galactic object SS433 is a high-mass X-ray binary at an advanced evolutionary stage (see the recent review \citealt{2020NewAR..8901542C}). The optical component most likely overfills its Roche lobe producing outflows through the inner and outer Lagrangian points L$_1$ and L$_2$. The binary orbital period of SS433 is $P_\mathrm{b}\simeq 13^\mathrm{d}.082.$ The system harbours an optically bright supercritical accretion disc with relativistic jets around the relativistic object
\citep{2004ASPRv..12....1F}. The disc precesses with a period of $P_\mathrm{prec}\simeq 162^\mathrm{d}.3$. The jets also nutate with a period of $P_\mathrm{nut}\simeq 6^\mathrm{d}.29.$ The parameters of the kinematic model of SS433 are found to be stable, on average, over the last 40 years \citep{2018ARep...62..747C}.
\cite{2007A&A...474..903B}, \cite{2008ARep...52..487D} and \cite{2018ARep...62..747C} detected traces of the orbital periodicity in the mass outflow velocity $\beta=v/c$ in the relativistic jets of SS433. These variations can be interpreted in terms of a possible small eccentricity of the binary orbit of SS433. Other indirect evidences of a non-zero binary eccentricity have also been found in \citet{2009ApJ...698L..23D} and \citet{2009MNRAS.397..849P}.

The stability of the system's parameters over decades looks enigmatic, considering a high mass-loss rate from the optical star and the binary system. The analysis of the limits on the orbital period variations in SS433 \citep{2018MNRAS.479.4844C, 2019MNRAS.485.2638C} suggests that the orbital period change strongly depends on the binary mass ratio  $q=M_\mathrm{X}/M_\mathrm{V}$ and the mode of the mass loss from the system. However, the available data for SS433 were insufficient to detect the orbital period change. 

In this paper, we examine the photometric orbital light curves of SS433 observed at the precession phases corresponding to the maximum disc opening. Our analysis revealed an asymmetric shape of the mean orbital light curve suggesting an elliptical orbit with an eccentricity of $e= 0.05\pm 0.01$. The
slaved accretion disc model in SS433 is more consistent with the non-zero binary eccentricity.
From the analysis of the observed light curves over 40 years, we have also found a strong evidence for a secular increase in the binary orbital period of SS433 at a rate of $\dot P_\mathrm{b}=(1.0\pm 0.3)\times 10^{-7}$~s~s$^{-1}$. Using our earlier results \citep{2018MNRAS.479.4844C, 2019MNRAS.485.2638C}, the detected binary orbital period change enabled us to put a lower limit on the binary mass ratio $q=M_\mathrm{X}/M_\mathrm{V}\gtrsim0.8$. This improved estimate strongly suggests that the relativistic object on SS433 is a black hole. 

Our findings support the evolutionary status of SS433 as a binary system in which the common envelope has not been formed. The system evolves as a semi-detached binary with a supercritical accretion disc around a black hole \citep{2017MNRAS.471.4256V}. 

\section{Observational data}

We have used photometric observations of SS433 in the V-filter carried out in 1978-2012 by different authors\footnote{See the compilation public database at \url{ftp://cdsarc.u-strasbg.fr/pub/cats/J/other/PZ/31.5/} by V. Goranskij}. These data were completed by our BVRcIc photometric observations carried out in 2019-2020 with the automatic  telescope RC600 of the Caucasian Mountain Observatory of Sternberg Astronomical Institute \citep{Berdnikov2020}. 
A more detailed discussion of our photometric and spectral monitoring of SS433 with CMO SAI MSU telescopes will be presented in a separate work. We also filled the missed interval 2012-2018 with photoelectric AAVSO data.

The differential photometry relative to a  reference star has been performed. The finding chart of SS433 and the reference stars can be found in the catalog of highly evolved close binaries \citep{1996hecb.conf.....C}. In total, in 2019-2020, V-photometric observations of SS433 were carried out during 276 nights. They were utilized to construct the precessional and orbital light curves of SS433. In our analysis, we have used the orbital light curves constructed from all available photometric data close to moment T$_3$ of the precessional period. At these moments, corresponding to a maximum separation of the moving emission lines of SS433, the accretion disc is maximally open to the observer's view. The data were selected within the time intervals $T_3\pm 0.2P_\mathrm{prec}$. 

\section{Analysis of the orbital light curve of SS433}

\begin{figure}
	\includegraphics[width=\columnwidth]{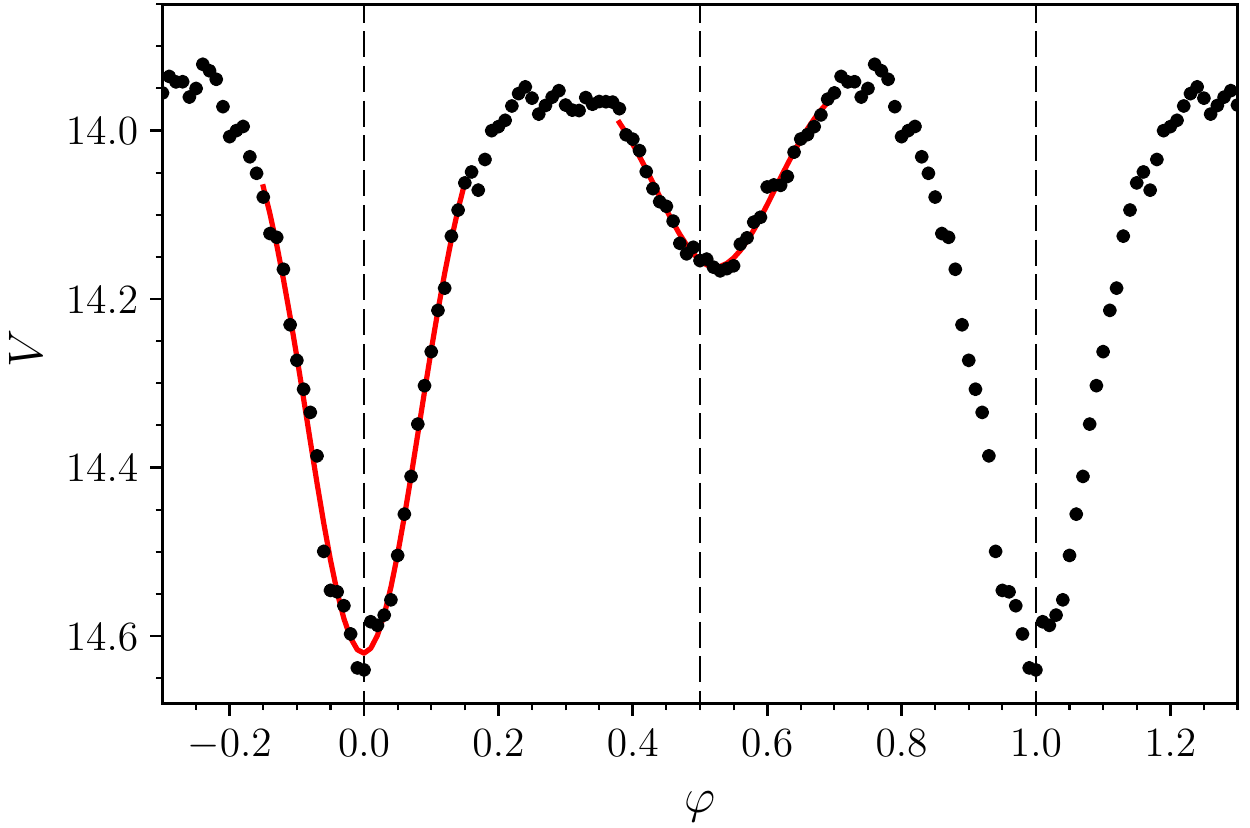}
    \caption{The average orbital light curve (black dots) constructed using 909 individual measurements taken in 1978-2012 inside the precession phase interval $T_3\pm0.2P_\mathrm{prec}.$ Red curves approximate the minima by Gaussians (see text).
     }
    \label{f:pc}
\end{figure}

The examination of the new photometric data of SS433 revealed a delay in the occurrence of the primary minima of the orbital light curve compared to the orbital ephemeris derived by V. Goranskii from 1978-2012 data \citep{2011PZ.....31....5G}, suggesting a possible change in the orbital period of SS433. In order to measure the phase delay precisely, we  reanalyzed all data in one manner.
First of all, we have rectified the entire light curve, removed peculiar features (flares, dips, etc.) and combined close in time ($<1^h$) observations. Then, we 
have calculated the mean orbital curve $\overline V(\varphi)$ (see Fig. \ref{f:pc}) using 1978-2012 data (which did not show signs of the change in the orbital period) falling within the precession phase intervals $T_3\pm 0.2P_\mathrm{prec},$ when the accretion disc is maximally open to the observer and irregular variations of the light curve are minimal. The mean light curve as a function of the orbital phase $\varphi$ was calculated by averaging the data 
$\overline{V}(\varphi) = \int{V(\phi)g(\varphi-\phi)}{\rm d}\phi$ by a Gaussian profile with a width of $\sigma_{\varphi}=0.03.$ The points deviated by more than $3\sigma$ from the mean value were removed, after which the procedure is repeated. 

The full time interval of observations was subdivided in eight approximately equal $\sim 5$-years' segments. Phase delays inside each time segments were calculated in two ways:

\textit{Method 1}. For each interval, the template $\overline V(\varphi)$ was fitted to the photometric data $V(\varphi)$ around $T_3\pm0.2P_{\rm prec}$ by adjusting three free parameters $a$, $\Delta \varphi$, $c:$
$V(\varphi) = a\overline V(\varphi-\Delta \varphi)+c$ (see Appendix\,\ref{appA} for the individual phase curves). 
The same procedure but applied only for the primary minimum $-0.25<\varphi<0.25$ yields essentially the same result. 

\textit{Method 2}. Instead of using the template $\overline V(\varphi),$ we approximated the primary minimum $-0.15<\varphi<0.15$ by a Gaussian with four free parameters. The Gaussian was adjusted to the data inside each interval.

\begin{figure}
	\includegraphics[width=\columnwidth]{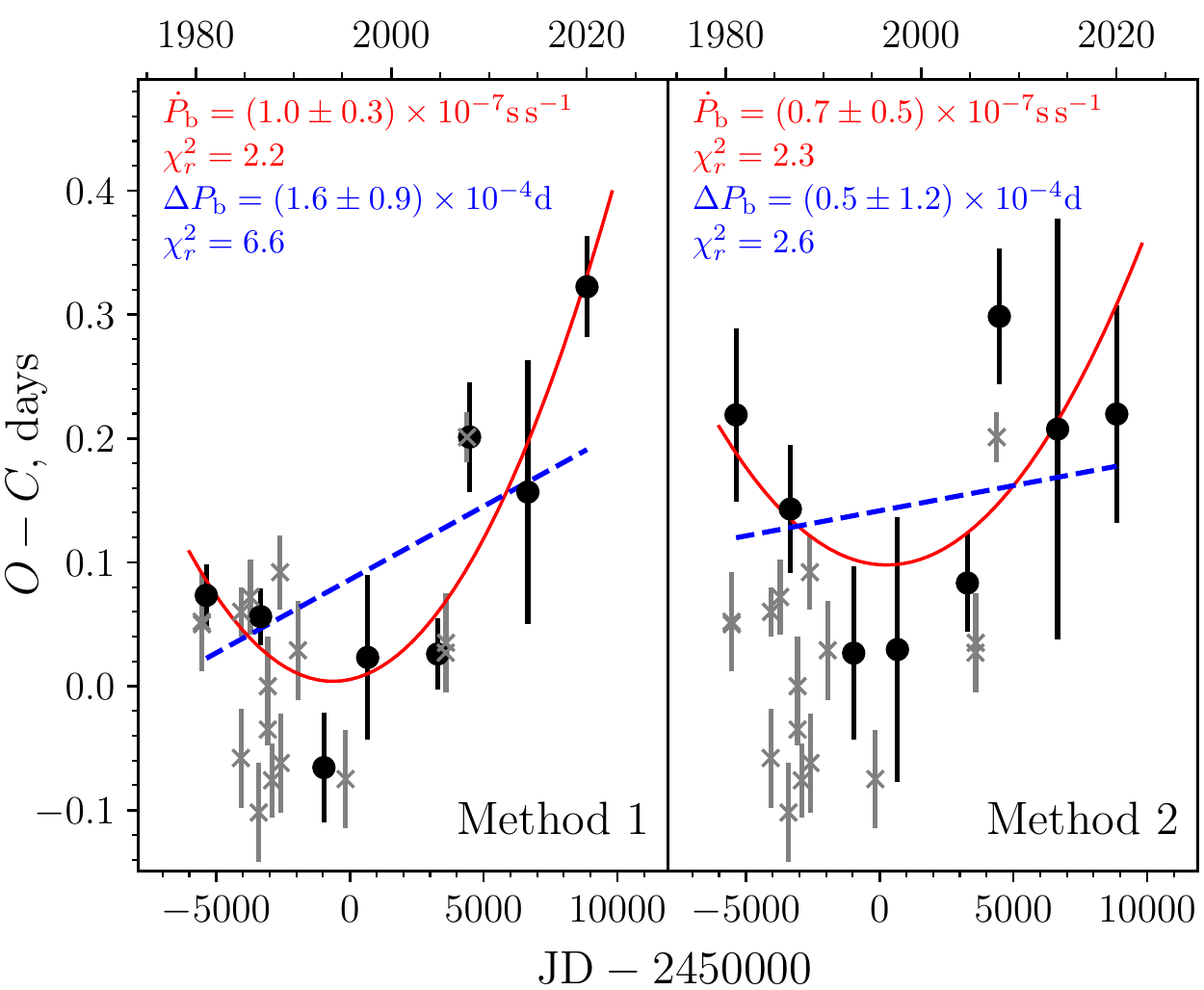}
    \caption{$O-C$ residuals of SS433 relative to the ephemeris with a constant orbital period of $P_\mathrm{b}=13^\mathrm{d}.08223$ calculated by various methods. Filled circles with error bars are obtained in the present paper.
    Grey crosses are for $O-C$ residuals for data from the literature. Red solid curves show a parabolic fit, blue dashed lines correspond to a linear fit. 
    The orbital period derivative $\dot{P}_\mathrm{b}$ 
    and the correction to the constant orbital period $\Delta P_\mathrm{b}$ are shown for each case.
    }
    \label{f:oc}
\end{figure}

In order to study variations of phase in the cycle of periodicity, it is
convenient to express the phase delays in terms of `Observed minus Calculated' residuals $(O-C)_{1,2}$ \citep{2005ASPC..335....3S}, where the index means the method used. The obtained $(O-C)_{1,2}$ residuals are shown in Fig. \ref{f:oc}. For comparison, in this Figure we present the most accurate $(O-C)_{\rm L}$ residuals with an error of less than $0^\mathrm{d}.05$ calculated from the published primary minima moments computed with the same constant orbital period (see Table 1 in \citealt{2011PZ.....31....5G}). It is seen that $(O-C)_1$ residuals are in good agreement with $(O-C)_{\rm L}$ and have smaller errors than $(O-C)_2.$ The values of $(O-C)_{1},$ as well as $(O-C)_{L}$ taken from the literature, are better described by a quadratic than by a linear fit.  The orbital period change rate  $\dot{P}_\mathrm{b}$ for the quadratic fit, the correction to the ephemeris $\Delta P_\mathrm{b}$ for the linear fit, and the corresponding values of the reduced $\chi^2_r$ are shown in Fig. \ref{f:oc}.
Note that the quadratic approximation for $(O-C)_{2}$ residuals yields close $\dot{P}_\mathrm{b}$ as for $(O-C)_{1}$ but due to large errors allows both quadratic and linear fits.

To summarize, Method 1 appears to be more reliable and gives the results consistent with previous estimates. Therefore, we will adopt the rate of the orbital binary period change of SS433
$\dot{P}_\mathrm{b}=(1.0\pm0.3)\times10^{-7}$s~s$^{-1}$.  
The new ephemeris for the primary orbital minima in SS433 reads:
\begin{center}
\begin{tabular}{crrl}
 $T_{\rm min}=$ &2451737.54 &  $+13.08246E$       &$+6\times10^{-7}E^2.$ \\
               &  $\pm$.04 &   $\pm$.00005\,~\,\,&  $\pm$2 
\end{tabular}
\end{center}
 The Julian dates of the primary minima of SS433 in each five-years' data intervals calculated from $(O-C)_1$ residuals are listed below: 
\\
\begin{tabular}{ccc}
2444633.94$\pm$0.03,&2446661.67$\pm$0.02,&2449029.43$\pm$0.04,\\
2450664.80$\pm$0.07,&2453281.25$\pm$0.03,&2454471.91$\pm$0.04,\\
2456656.59$\pm$0.11,&2458867.66$\pm$0.04.& \\
    \end{tabular}



\section{Orbital eccentricity of SS433}

Fig. \ref{f:pc} shows the average orbital light curve of SS433 (black dots) suggesting that the secondary minimum is shifted with respect to the phase 0.5. In order to quantify the shift, we have fitted both minima by Gaussians (the red curves in Fig. \ref{f:pc}) with the fixed out-of-eclipse level $V_0=13.93.$ The distance between the minima is found to be $\varphi_2-\varphi_1=0.5231\pm0.0012,$ the Gaussian width of the primary and secondary minima is $\sigma_1^* = 0.0834\pm 0.0009$ and $\sigma_2^* = 0.0885\pm0.0014,$ respectively, the depth of the primary and the secondary minimum is $0.694\pm0.006$ and  $0.235\pm0.003$ magnitudes, respectively. Measured values of $\sigma^*_{1,2}$ should be corrected for introduced smoothing $\sigma_\varphi=0.03$ as $\sigma^2_{1,2}=\sigma^{*2}_{1,2}-\sigma^2_{\varphi}.$ The difference in the position of the minima and their width enables determining  the orbital eccentricity $e$ and  periastron longitude $\omega$ \citep{2009ebs..book.....K} 
\beq{e:ew}
\begin{cases}
e\cos\omega = \frac{\pi}{1+\csc^2i}\left(\varphi_2-\varphi_1-0.5\right),\\
e\sin\omega = \frac{\sigma_2-\sigma_1}{\sigma_2+\sigma_1},
\end{cases}
\eeq
where $i=79\degr$ is the orbital inclination. The solution of these two equations yields  $e=0.047\pm0.007,$ $\omega=40\degr\pm20\degr.$

To test these results, we calculated the phase difference $\varphi_2-\varphi_1$ and width of the eclipses by other methods. A parabolic fit to the minima $a(\varphi-\varphi_{1,2})^2+C$ with the width defined as $2\left[0.5(V_0-C)/a\right]^{0.5}$ gives $e=0.055\pm0.015$ and $\omega=50\pm30\degr$. The calculation of the centre of gravity for each minima and with the width measured via the second moment $\sigma^{*2}_{1,2}=\sum_i(\varphi_i-\varphi_{1,2})^2(\overline{V}_i-V_0)/\sum_i(\overline{V}_i-V_0)$ leads to $e=0.05$ and $\omega=30\degr$ (errors are undefined). The 0.01$^\mathrm{m}$ uncertainty in the assumed value of $V_0$ changes the parameters insignificantly, but we can include it as an additional free parameter. Searching for a single $V_0$ for both Gaussians in our fitting procedure leads to $e=0.039\pm0.012$ and $\omega=20\pm60\degr.$  Fitting each minima with a Gaussian curve with individual constants $V_0$ leads to $e=0.065\pm0.035$ and $\omega=60\pm70\degr$. Assuming the sum of a linear slope and a Gaussian curve gives an even higher value of $e=0.10\pm0.05$ and $\omega\sim75\degr.$ A major source of uncertainties is related to the eclipse width measurements, which impacts mostly $\omega$. By considering only the position of the minima, we can set a lower limit on $e,$ which in all cases is $e>0.03.$
These tests allow us to conclude that a binary eccentricity of $0.05\pm0.01$ is consistent with all methods we tried.

\section{Discussion}

The detection of a non-zero orbital eccentricity and evidence for the binary period increase in SS433 have important implications for the evolutionary status of this high-mass binary system. 
Before considering them, we should discuss the robustness of our findings. First of all, the orbital eccentricity should be accompanied by the apsidal line motion, and second, the presence of a third body can mimic the deviations of the $O-C$ residuals from a linear fit.

\subsection{Apsidal advance}

The periastron advance of an elliptical binary orbit may affect the orbital period increase derived from the $O-C$ diagram. The accuracy of the orbital light curves constructed within the five-years' time intervals is insufficient to calculate $e$ and $\omega$ using the five-years' data only. Therefore, we have subdivided all available data into two times intervals, spanning 1978-2000 and 2001-2020, and determined the eccentricity and average periastron longitudes for each of them:
 $e=0.04\pm0.01,$ $\omega_{1990}=40\pm20\degr,$ $\omega_{2010}=0\pm10\degr.$

Thus, over four decades, the change in $\omega$ could be   $\sim90\degr,$ which is consistent with the expected apsidal motion period $U\sim 150$ years for the apsidal motion constant $k_2\sim 2\times 10^{-3}$
(see, for example, calculations of $k_2$ for massive evolved $15-20 {\rm M}_\odot$ stars in   \citealt{2016A&A...594A..33R}). As the 
amplitude of the apsidal motion is 
$A=eP_\mathrm{b}/2\pi \approx 0^\mathrm{d}.1,$ in a quarter of apsidal motion period the deviation of the sine line from the straight line in the $O-C$ residuals would be less than $0.15A \approx 0^\mathrm{d}.02,$ which does not affect the $O-C$ residuals and the derived estimate of $\dot{P}_\mathrm{b}$ significantly.


\subsection{A third body hypothesis}

The presence of a third body in SS433 
may also lead to deviations of the $O-C$ residuals from a linear fit due to the motion of the binary system around the barycenter in 
the possible triple system. For a circular orbit of the hypothetical third body around a binary system, deviations of the binary's $O-C$ residuals from the linear law read \citep{1973bmss.book.....B}:
\beq{e:tb}
\Delta_{O-C} = \frac{a^\prime\sin i^\prime}{c}\sin\left[\frac{2\pi(t-T_0^\prime)}{P^\prime}\right],
\eeq
where $a^\prime,$ $i^\prime, $ $P^\prime,$ $T_0^\prime$ are the radius of the third body's orbit, its inclination, orbital period and the initial epoch, respectively. 
\begin{figure}
	\includegraphics[width=\columnwidth]{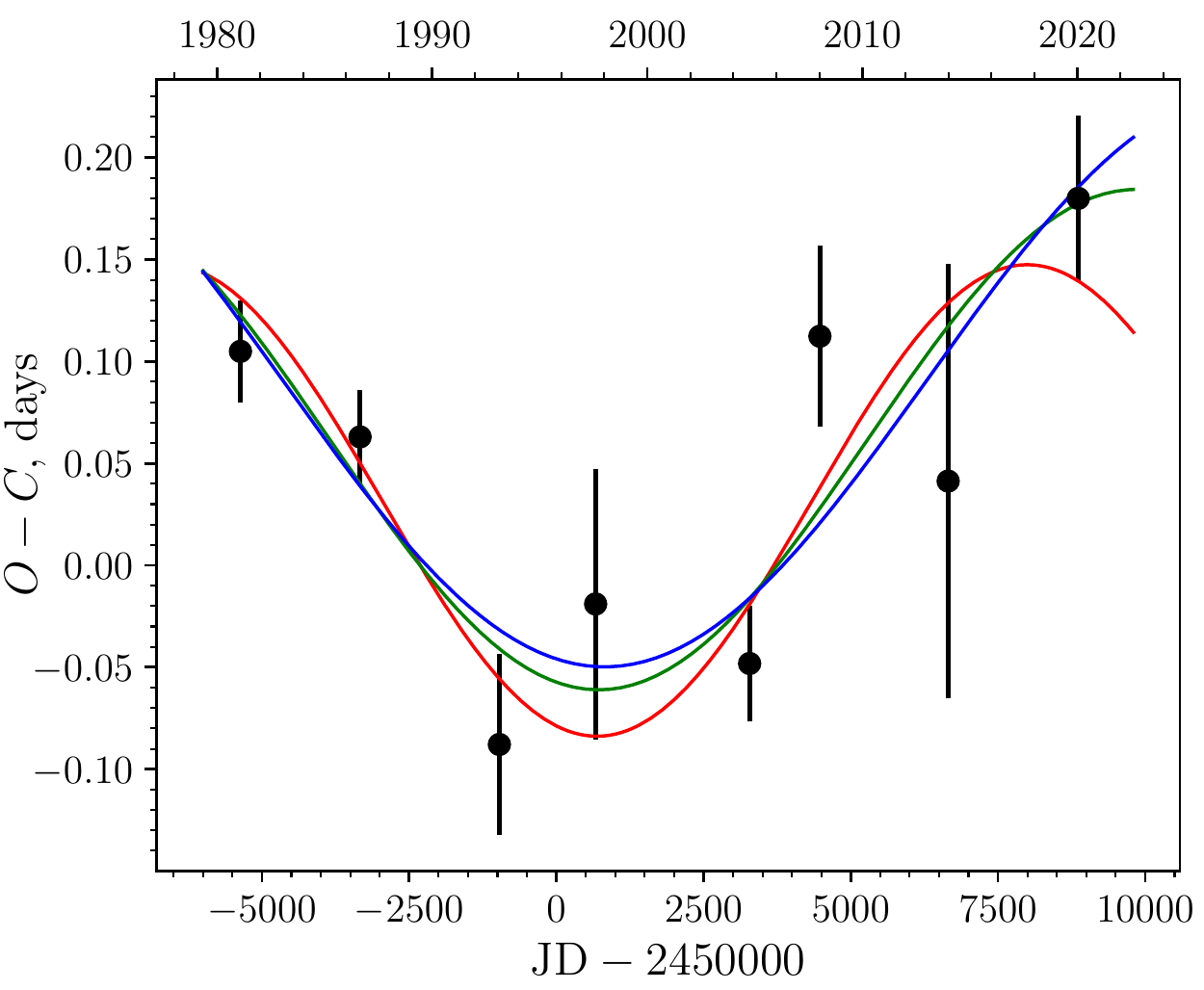}
    \caption{ Black dots: the $O-C$ residuals with subtracted linear trend corresponding to $\Delta P_\mathrm{b} =1.6\times10^{-4}$ d (see Fig.~\ref{f:oc}). Red, green, and blue lines approximate the residuals by a sine law with $P^{\prime}=40,$ 50, 60 years. $\chi^2_r=1.7,$ 1.7, 1.8, respectively.}
    \label{f:ocsin}
\end{figure}
These parameters can constrain the third body's mass:
\beq{e:m3}
\frac{(M_3\sin i^\prime)^3}{(M_{\rm X}+M_{\rm V}+M_3)^2}=\frac{(a^{\prime}\sin i^{\prime})^3}{P^{\prime 2}}\,.
\eeq
Here the orbital period is in years, mass in the solar units, and the orbital semi-axis in astronomical units.

In order to find the smallest $(a^{\prime}\sin i^{\prime})^3/P^{\prime2}$ from $O-C$, we
subtract the linear component (the dashed blue line in Fig.\ref{f:oc}) from the observed values of $(O-C)_1$ and approximate the residuals by a sine law for a set of trial periods $P^{\prime}>40$ years. The minimal ratio $(a^{\prime}\sin i^{\prime})^3/P^{\prime2}\approx4$ is found at $P^{\prime}\approx50$ years.

Using the low limit on the mass ratio $M_{\rm X}/M_{\rm V}>0.6$ \citep{2019MNRAS.485.2638C} and
$M_{\rm V}=10-15{\rm M}_\odot$ , we get for
the third star's mass $M_3>16-20{\rm M}_\odot.$ A star with mass exceeding that of the donor star should manifest itself in the optical spectra.
However, the high-resolution optical spectra of SS433 \citep{2004ApJ...615..422H, 2008ApJ...676L..37H} that were used to measure the orbital radial velocity of the donor star did not show traces of additional stationary lines from the third star. In addition, the presence of the third star with a mass exceeding that of the donor star in SS433 would be strange from the evolutionary point of view. The optical star in SS433 has already left the main sequence, therefore the third star with higher mass would appear as a 
post-main sequence bright blue (super)giant or even as a stellar remnant, only as a black hole in the latter case. 

Thus, we conclude that although we could not fully reject the hypothesis of the third star causing the observed $O-C$ deviations from the linear law, it seems to be rather unlikely.  Future long-term measurements of the orbital period of SS433 are needed to 
check this hypothesis. These observations are continuing at the CMO SAI.

\subsection{Binary period change in SS433}

The detected secular increase in the orbital binary period  of SS433 $\dot P_\mathrm{b}=(1.0\pm 0.3)\times 10^{-7}$~s~s$^{-1}$ has important implications for the binary system's parameters. Of these, the major is an improved estimate of the binary mass ratio $q=M_\mathrm{X}/M_\mathrm{V}$. In order to derive it, consider the model of a binary system with isotropic wind from a supercritical accretion disc around the relativistic compact object (the Jeans mode) and a possible mass-loss from the system with the dimensionless specific angular momentum as inferred from observations of the external shell outflow in SS433 (see \citealt{2019MNRAS.485.2638C} for more detail). In this model, the fractional change in the binary orbital period reads\footnote{There is a misprint in the caption to Fig. 1 in \citep{2019MNRAS.485.2638C}: the quadratic equation in the case of the mass outflow via Jeans mode only with constant orbital period should read $q^2+(2/3)q-1=0$}:
\beq{e:q}
\frac{\dot P_\mathrm{b}}{P_\mathrm{b}}=-\frac{\dot M_\mathrm{V}}{M_\mathrm{V}}\frac{3q^2+2q-3\beta -3K(1-\beta)(1+q)^{5/3}}{q(1+q)}\,.
\eeq
Here, the mass-loss rate from the optical star with the mass $M_\mathrm{V}$ is $\dot M_\mathrm{V}$, the mass accreted by the compact object is assumed to be zero, the mass outflow from the supercritical accretion disc via the Jeans mode (i.e. with the specific orbital angular momentum of the accretor) is $\beta\dot M_\mathrm{V}$, $\beta\le 1$, and the mass outflow from the system via a circumbinary shell is $(1-\beta)\dot M_\mathrm{v}$.
The dimensionless coefficient $K$ specifying the angular momentum loss via the circumbinary disc is assumed to be $K=4.7$, given the new \textit{Gaia} DR2 distance to SS433 $d\approx 3.8$~kpc \citep{2021MNRAS.502.5455A} and $M_\mathrm{V}=10 M_\odot$ instead of  $d=5.5$~kpc and $M_\mathrm{V}=15 M_\odot$ used in  \cite{2019MNRAS.485.2638C}.

It is convenient to recast \Eq{e:q} to the form:
\beq{e:q1}
(3-A)q^2+(2-A)q-3\beta-3K(1-\beta)(1+q)^{5/3}=0
\eeq
where the dimensionless parameter $A$ is
\[
A\equiv-\myfrac{\dot P_\mathrm{b}}{P_\mathrm{b}}\myfrac{M_\mathrm{V}}{\dot M_\mathrm{V}}\,.
\]
With fixed $\dot P_\mathrm{b}/P_\mathrm{b}$, the parameter $A$ is determined by the mass-loss rate from the optical star and its mass. 
It is easy to see that \Eq{e:q1} has no real solutions for $A\ge 3$. This implies that for the minimum value of ${\dot P}_\mathrm{b}=0.7\times 10^{-7} $~s~s$^{-1}$, the mass-loss rate form the optical star cannot be lower than $\sim 7\times 10^{-6} {\rm M}_\odot\mathrm{yr}^{-1}$. 

\begin{figure}
	\includegraphics[width=\columnwidth]{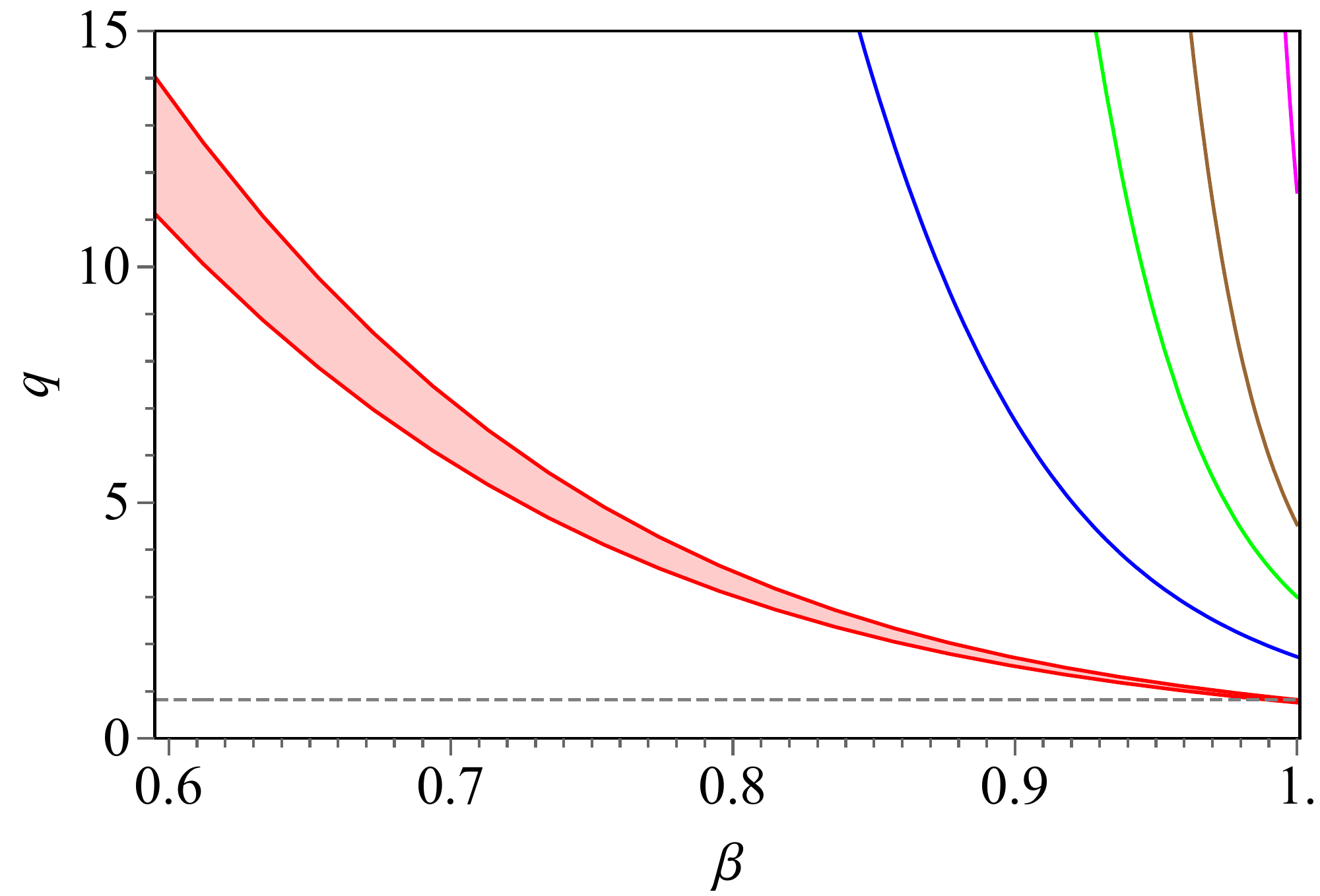}
    \caption{Constraints on the mass ratio of SS433 for the fiducial mass-loss rate $\dot M_\mathrm{V}=10^{-4}{\rm M}_\odot$yr$^{-1}$ from the optical star with mass $M_\mathrm{V}=10 {\rm M}_\odot$ and
    the dimensionless parameter $K=4.7$. 
    The shaded band corresponds to the $\pm 1\sigma$ errors in the orbital period derivative. Solid blue, green, brown and magenta curves show the solution of \Eq{e:q1} for the parameter $A=2, 2.5, 2.7$ and 2.9, respectively.
    The horizontal line shows the solution of \Eq{e:q} $q_\mathrm{min}\simeq 0.82$ corresponding to the observed $\dot{P}_\mathrm{b}=1.0\times 10^{-7}$~s~s$^{-1}$ for $\beta=1$ (only Jeans mode mass outflow).}
    \label{f:f1}
\end{figure}

Fig. \ref{f:f1} presents solutions $q(\beta)$ of \Eq{e:q1} for different values of the parameter $A$. The shadowed strip corresponds to the uncertainty in the measured value of ${\dot P}_\mathrm{b}$ with fixed $\dot M_\mathrm{V}=10^{-4} {\rm M}_\odot\mathrm{yr}^{-1}$ and $M_\mathrm{V}=10 {\rm M}_\odot$. The lines above this strip are computed for $A=2, 2.5, 2.7$ and 2.9, respectively. The horizontal line corresponds to the solution of \Eq{e:q1} $q_\mathrm{min}\approx 0.82$ for $\beta=1$ (no mass outflow from the system through the external Lagrangian point). It is seen that the decrease in the mass-loss rate of the optical star\footnote{More precisely, in the ratio $\myfrac{\dot M_\mathrm{V}}{ M_\mathrm{V}}$ that determines the parameter $A$.} increases the minimum possible binary mass ratio.


\section{Conclusions}

The analysis of long-term photometric observations of SS433, including the recent observations carried out at CMO SAI MSU (2019-2020), enabled us to discover a secular increase in the orbital period 
$\dot P_\mathrm{b}=(1.0\pm0.3)\times 10^{-7}$~s~s$^{-1}$. Using the physically motivated model of the isotropic mass re-emission (the Jeans mode) from a supercritical accretion disc and the mass loss from the system via the external Lagrangian point L$_2$ \citep{2018MNRAS.479.4844C,2019MNRAS.485.2638C}, we have improved the estimate of the binary components' mass ratio, $q=M_\mathrm{X}/M_\mathrm{V}\gtrsim 0.8$. The detected increase in the orbital period also suggests the lower limit on the physically acceptable mass-loss rate from the optical star $|\dot M_\mathrm{V}|>7\times 10^{-6} {\rm M}_\odot\mathrm{yr}^{-1}$. For the optical star mass $M_\mathrm{V}=10 {\rm M}_\odot$ \citep{2011PZ.....31....5G}, the mass of the relativistic object in SS433 is $M_\mathrm{X}\gtrsim 8 {\rm M}_\odot$, i.e. it is a black hole with a mass close to the 'standard' mass of black holes in Galactic high-mass X-ray binaries (see, for example, \citealt{2016PhyU...59..910C}). A recent update to SS433's distance \citep{2021MNRAS.502.5455A} can reduce the optical star mass by $\approx20$ per cents, but leaves the mass ratio estimates from our analysis unchanged.
We stress that the mass ratio lower limit $q>0.8$, obtained from the long-term optical observations of SS433, is in agreement with an independent estimate $q\gtrsim 0.4\div 0.8$ derived from the analysis of X-ray observations of S433 by the INTEGRAL satellite \citep{2020NewAR..8901542C}. This limit was obtained in the model in which the optical star overfills its Roche lobe and the mass flow occurs through L$_1$ and L$_2$ points. 

Note that \Eq{e:q} suggests that a decrease in the mass ratio $q$ leads to the sign change of the period derivative. For example, in the limiting case $\beta=1$, this occurs at $q\lesssim 0.72$. Therefore, for low mass ratios in SS433, we would expect a \textit{decrease} in the orbital period. Thus, the detected increase in the orbital period of SS433 implies $q=M_\mathrm{X}/M_\mathrm{V}>0.72$. For the $10 {\rm M}_\odot$ optical companion, this firmly excludes a neutron star as the relativistic object in SS433.

The form of the optical eclipsing light curve of SS433, constructed from long-term photometrical observations at the phases near the maximum opening of the precessing accretion disc, $T_3\pm 0.2P_{\rm prec}$, enabled us to reveal an ellipticity of the SS433 orbit. The shift of the secondary minimum relative to the middle between two consecutive primary minima puts a lower limit for the orbital eccentricity of SS433 $e>0.03,$ while the fact that the secondary minimum is wider than the primary one allowed us to find the value of $e=0.05\pm0.01$ and the periastron longitude $\omega=40^\circ$ at epoch $\sim 1990$.

As noted earlier \citep{2018ARep...62..747C}, the orbital ellipticity in SS433 can be naturally explained in the model of the slaved disc precession, where the disc traces the precession of the optical star's rotational axis. Indeed, the fact that the rotational axis of the optical star in SS433 is misaligned with the orbital momentum (which may be a consequence of an asymmetric core collapse leading to the formation of the relativistic component, see \citealt{1974ApJ...187..575R,1981SvAL....7..401C}) implies an asynchronous rotation of the star with the orbital period. Meanwhile, the theory of tidal synchronization of stars in binaries due to dissipation of the orbital energy in dynamical tides \citep{1977A&A....57..383Z,1989A&A...220..112Z} suggests that the synchronization of the axial and orbital rotation of the star in a close binary occurs earlier than the tidal circularization of the orbit. As the optical star in SS433 is not synchronized with the orbital period, the orbit in this binary system can be eccentric, which is actually observed. Our discovery of the orbital ellipticity of SS433 is consistent with the slaved disc model that requires optical star's rotational axis misalignment with the orbital angular momentum. The average stability of parameters of the kinematic model of SS433 observed over more than four decades \citep{2018ARep...62..747C} also supports the slaved disc model.

\section*{Acknowledgements}

We thank the referee, Dr. S. Perez, for valuable comments.
We thank Dr. N. Ikonnikova and M. Burlak for help in RC600 observations. 
The work of AMCh and AVD is supported by the RSF grant 17-12-01241 (analysis of the O-C light curve delay).
The work is supported by the Scientific and Educational School of M.V. Lomonosov Moscow State University
'Fundamental and applied space research'. The authors acknowledge support from M.V.Lomonosov Moscow State University Program of Development.

\section*{Data availability}
Photometric observations of SS433 before 2018 used in the present paper are available via CDS at \url{ftp://cdsarc.u-strasbg.fr/pub/cats/J/other/PZ/31.5} and
AAVSO at \url{https://www.aavso.org}. The CMO RC600 photometric observations of SS433 are  available online in the supplementary materials.




\bibliographystyle{mnras}
\bibliography{ss433} 
\appendix
\section{Orbital phase curves for each time interval}\label{appA}
\begin{figure}
	\includegraphics[width=\columnwidth]{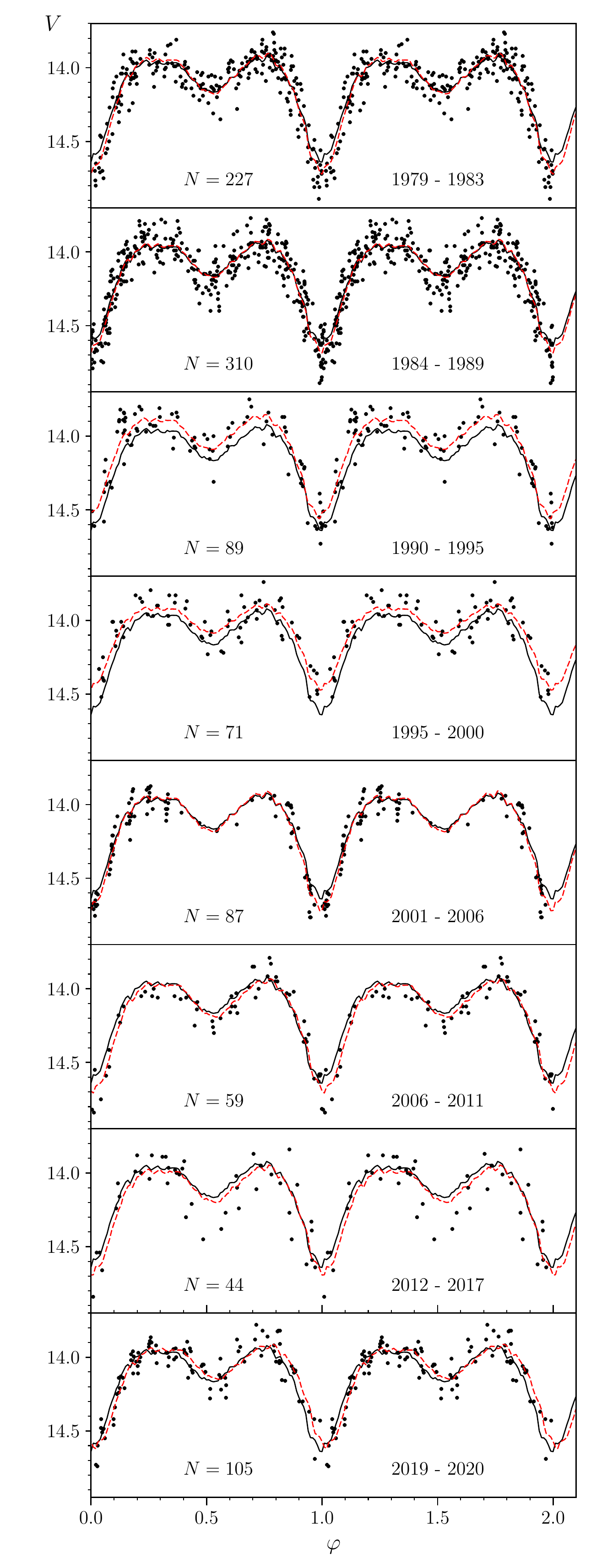}
    \caption{Observed light curves of SS433 phased with the orbital period for each time interval (dots). The solid black line is the orbital light curve template with the assumed constant orbital period. The red dashed lines are for the scaled and shifted light curves best-fitting the observations using Method 1. $N$ is the number of observational points used for the fitting.
     }
    \label{f:lc}
\end{figure}
\bsp	
\label{lastpage}
\end{document}